
\font\twelverm=cmr10 scaled 1200    \font\twelvei=cmmi10 scaled 1200
\font\twelvesy=cmsy10 scaled 1200   \font\twelveex=cmex10 scaled 1200
\font\twelvebf=cmbx10 scaled 1200   \font\twelvesl=cmsl10 scaled 1200
\font\twelvett=cmtt10 scaled 1200   \font\twelveit=cmti10 scaled 1200

\skewchar\twelvei='177   \skewchar\twelvesy='60

\def\twelvepoint{\normalbaselineskip=12.4pt
  \abovedisplayskip 12.4pt plus 3pt minus 9pt
  \belowdisplayskip 12.4pt plus 3pt minus 9pt
  \abovedisplayshortskip 0pt plus 3pt
  \belowdisplayshortskip 7.2pt plus 3pt minus 4pt
  \smallskipamount=3.6pt plus1.2pt minus1.2pt
  \medskipamount=7.2pt plus2.4pt minus2.4pt
  \bigskipamount=14.4pt plus4.8pt minus4.8pt
  \def\rm{\fam0\twelverm}          \def\it{\fam\itfam\twelveit}%
  \def\sl{\fam\slfam\twelvesl}     \def\bf{\fam\bffam\twelvebf}%
  \def\mit{\fam 1}                 \def\cal{\fam 2}%
  \def\tt{\twelvett}
  \textfont0=\twelverm   \scriptfont0=\tenrm   \scriptscriptfont0=\sevenrm
  \textfont1=\twelvei    \scriptfont1=\teni    \scriptscriptfont1=\seveni
  \textfont2=\twelvesy   \scriptfont2=\tensy   \scriptscriptfont2=\sevensy
  \textfont3=\twelveex   \scriptfont3=\twelveex  \scriptscriptfont3=\twelveex
  \textfont\itfam=\twelveit
  \textfont\slfam=\twelvesl
  \textfont\bffam=\twelvebf \scriptfont\bffam=\tenbf
  \scriptscriptfont\bffam=\sevenbf
  \normalbaselines\rm}

\def\beginlinemode{\endmode
  \begingroup\parskip=0pt \obeylines\def\\{\par}\def\endmode{\par\endgroup}}
\def\beginparmode{\endmode
  \begingroup \def\endmode{\par\endgroup}}
\let\endmode=\par
{\obeylines\gdef\
{}}
\def\singlespace{\baselineskip=\normalbaselineskip}

\def\oneandahalfspace{\baselineskip=\normalbaselineskip
  \multiply\baselineskip by 3 \divide\baselineskip by 2}
\def\doublespace{\baselineskip=\normalbaselineskip \multiply\baselineskip by 2}

\newcount\firstpageno
\firstpageno=2
\footline={\ifnum\pageno<\firstpageno{\hfil}\else{\hfil\twelverm\folio\hfil}\fi}
\let\rawfootnote=\footnote              
\def\footnote#1#2{{\rm\singlespace\parindent=0pt\rawfootnote{#1}{#2}}}
\def\raggedcenter{\leftskip=4em plus 12em \rightskip=\leftskip
  \parindent=0pt \parfillskip=0pt \spaceskip=.3333em \xspaceskip=.5em
  \pretolerance=9999 \tolerance=9999
  \hyphenpenalty=9999 \exhyphenpenalty=9999 }
\def\dateline{\rightline{\ifcase\month\or
  January\or February\or March\or April\or May\or June\or
  July\or August\or September\or October\or November\or December\fi
  \space\number\year}}
\def\received{\vskip 3pt plus 0.2fill
 \centerline{\sl (Received\space\ifcase\month\or
  January\or February\or March\or April\or May\or June\or
  July\or August\or September\or October\or November\or December\fi
  \qquad, \number\year)}}
\hsize=6.5truein
\hoffset=1truein
\vsize=8.9truein
\voffset=1truein
\parskip=\medskipamount
\twelvepoint            
\doublespace            
\overfullrule=0pt       
\def\preprintno#1{
 \rightline{\rm #1}}    

\def\title                      
  {\null\vskip 3pt plus 0.2fill
   \beginlinemode \doublespace \raggedcenter \bf}

\def\author                     
  {\vskip 3pt plus 0.2fill \beginlinemode
   \singlespace \raggedcenter}

\def\affil                      
  {\vskip 3pt plus 0.1fill \beginlinemode
   \oneandahalfspace \raggedcenter \sl}

\def\abstract                   
  {\vskip 3pt plus 0.3fill \beginparmode
   \doublespace \narrower ABSTRACT: }

\def\endtitlepage               
  {\endpage                     
   \body}

\def\body                       
  {\beginparmode}               

\def\head#1{                    
  \filbreak\vskip 0.5truein     
  {\immediate\write16{#1}
   \raggedcenter \uppercase{#1}\par}
   \nobreak\vskip 0.25truein\nobreak}

\def\refto#1{$|{#1}$}           

\def\references                 
  {\head{References}            
   \beginparmode
   \frenchspacing \parindent=0pt \leftskip=1truecm
   \parskip=8pt plus 3pt \everypar{\hangindent=\parindent}}

\gdef\refis#1{\indent\hbox to 0pt{\hss#1.~}}    

\gdef\journal#1, #2, #3, 1#4#5#6{               
    {\sl #1~}{\bf #2}, #3, (1#4#5#6)}           

\def\refstylenp{                
  \gdef\refto##1{ [##1]}                                
  \gdef\refis##1{\indent\hbox to 0pt{\hss##1)~}}        
  \gdef\journal##1, ##2, ##3, ##4 {                     
     {\sl ##1~}{\bf ##2~}(##3) ##4 }}

\def\refstyleprnp{              
  \gdef\refto##1{ [##1]}                                
  \gdef\refis##1{\indent\hbox to 0pt{\hss##1)~}}        
  \gdef\journal##1, ##2, ##3, 1##4##5##6{               
    {\sl ##1~}{\bf ##2~}(1##4##5##6) ##3}}

\def\endreferences{\body}

\def\figurecaptions             
  {\endpage
   \beginparmode
   \head{Figure Captions}
}

\def\endpage                    
  {\vfill\eject}
\def\endpaper                   
  {\endmode\vfill\supereject}

\def\ref#1{Ref. #1}                     

\def\frac#1#2{{\textstyle{#1 \over #2}}}
\def\half{{\textstyle{ 1\over 2}}}
\def\leaderfill{\leaders\hbox to 1em{\hss.\hss}\hfill}
\def\comp{{\rm C}\llap{\vrule height7.1pt width1pt depth-.4pt\phantom t}}
\def\fint{\rlap{$\biggl\rfloor$}\biggl\lceil}
\def\m@th{\mathsurround=0pt }
\singlespace
\preprintno{UFIFT-HEP-90-22}
\preprintno{VPI-IHEP-91/5}
\preprintno{Revised Version}
\preprintno{November 1991}

\title Nonlocal Yang-Mills

\author Gary Kleppe
\affil
Department of Physics,
Virginia Polytechnic Institute,
Blacksburg VA 24061
\vskip .5cm
\centerline{and}
\vskip .5cm
\author R. P. Woodard
\affil
Department of Physics
University of Florida
Gainesville, FL 32611

\abstract We present a very simple and explicit procedure for nonlocalizing the
action of any theory which can be formulated perturbatively. When the resulting
nonlocal field theory is quantized using the functional formalism --- with unit
measure factor --- its Green's functions are finite to all orders. The
construction also ensures perturbative unitarity to all orders for scalars with
nonderivative interactions, however, decoupling is lost at one loop when vector
and tensor quanta are present. Decoupling can be restored (again, to all
orders)
if a suitable measure factor exists. We compute the required measure factor for
pure Yang-Mills at order $g^2$ and then use it to evaluate the vacuum
polarization at one loop. A peculiar feature of our regularization scheme is
that the on-shell tree amplitudes are completely unaffected. This implies that
the nonlocal field theory can be viewed as a highly noncanonical quantization
of
the original, local field equations.

\endtitlepage
\doublespace

\centerline{\bf 1. Introduction}

At any point in the evolution of science there is always a ``simplest way'' to
perceive true results, yet insights of great power sometimes emerge from
adopting novel perspectives. The putative finiteness of superstring loop
amplitudes is currently best understood through the Polyakov representation
[1],
a formalism which makes frequent and essential use of the fact that the first
quantized dynamical variable is an extended object. Few would care to view
finiteness from the perspective of string field theory; indeed it is not even
simple to compute {\it tree} amplitudes in this way [2]. However when the
attempt is made a very curious fact emerges: {\it the finiteness of string
field
theory would follow trivially from the nonlocality of its interactions} [3]
{\it were it not for the infinite number of component fields.}\footnote{*}{We
have suppressed a necessary qualifier in the interest of rhetorical trenchancy.
There are, of course, many sorts of nonlocal vertices and not all of them serve
to quench loop divergences. The ultraviolet finiteness of string field theory
can be ascribed to the fact that its vertices include nonlocal convergence
factors of the form $\exp(-\alpha' p^2)$.} Since this last feature is a direct
consequence of the infinite number of coordinates needed to describe an
extended
object we see that the much touted property of ``stringyness'' is not only of
no
use in quenching divergences, it is actually the only thing which can threaten
the otherwise manifest finiteness of a suitably nonlocal field theory.

To this observation must be added the fact that most of the extra fields of
string theory serve no other purpose than to distort its phenomenology in often
untestable and sometimes unacceptable ways. So why do we not instead consider
nonlocal theories based on a finite number of fields? Until very recently the
answer would have been that strings provide the only perturbatively consistent
and nonlocal theories which contain interacting vector and tensor quanta.

In fact it was long believed impossible to construct perturbatively consistent,
purely {\it scalar} theories with nonlocal interactions, despite years of
heroic
effort [4]. The problem in this case was how to construct an operator formalism
through canonical quantization. By exploiting the functional formalism directly
Polchinski exhibited a satisfactory model in 1984 [5], and a procedure for
deriving the associated operator formalism was finally given (in a different
context) by Ja\'en, Llosa and Molina in 1986 [6]. Though highly significant,
these developments still do not suffice for the consistent incorporation of
vector and tensor quanta because nonlocal interactions disrupt the local gauge
symmetries which are used to reconcile Lorentz invariance and perturbative
unitarity.

A two-stage procedure for avoiding this problem was discovered in early 1990
[7]. In the first stage a gauge invariant action is nonlocalized and then
fortified with an infinite series of carefully chosen higher interactions.
These
new interactions endow the theory with a strange sort of nonlinear and nonlocal
gauge invariance which does the crucial job of reconciling Lorentz invariance
and perturbative unitarity at tree order. The second stage consists of finding
a measure factor to make the functional formalism invariant under the nonlocal
gauge symmetry without compromising perturbative unitarity.\footnote{*}{We
thank
A. Polychronakos for showing us that a measure factor can always be found which
restores invariance. The nontrivial thing is to find one whose interactions are
entire functions of the derivative operator. Otherwise perturbative unitarity
is
threatened.} If such a measure factor can be found then the resulting
functional
formalism defines a set of Green's functions which are ultraviolet finite and
Poincar\'e covariant to all orders, and which reduce to give perturbatively
unitary scattering amplitudes. For QED the first stage was worked out
explicitly
and a proof was given that the second stage could be carried out as well. For
Yang-Mills and gravity only an indirect argument was presented that the first
stage could be completed and no results were reported regarding the second
stage.

The purpose of this paper is to give a very simple procedure for constructing
nonlocal versions of Yang-Mills, gravity or any other local field theory which
has a perturbative formulation. This procedure, the work of section 2, has two
applications. First, it provides a regularization technique which can be used
in
any theory. Although the method is no more difficult to implement than
dimensional regularization it is also no easier; its selling point is that if a
suitable measure factor exists then the nonlocalized theory preserves
physically
equivalent versions of any and all continuous symmetries of its parent. This
fact is proven in appendix B. The second application for nonlocalization is the
generation of new fundamental theories. If a suitable measure factor exists
then
the nonlocalized theory appears to meet all the requirements for a fundamental
model and there is really no reason to take the local limit.

In section 3 we implement the second stage for Yang-Mills to the first
nontrivial order. The resulting Feynman rules are used in section 4 to compute
the one loop correction to the gluon propagator. Section 5 discusses the
possibility that nonlocalization produces consistent fundamental theories. To
do so the resulting models must avoid the nonperturbative problems which occur
in most nonlocal theories, including string field theory [8]. These problems
derive largely from the existence of higher derivative solutions to the
classical field equations; we will show in appendix A that nonlocalization does
not engender such solutions. This prompts us to advance a radical proposal for
extricating physics from the apparent impasse between gravitation and quantum
mechanics. Our conclusions comprise section 6.

Before proceeding we should comment on the distinction between nonlocalization
and two other invariant schemes with which it is sometimes confused. In the
{\it higher derivative} method one attempts to regulate the theory by adding
terms with higher covariant derivatives and/or higher powers of the field
strength. This makes the propagator fall off faster for large Euclidean
momenta but it also induces interactions with more factors of the momenta. The
two effects balance at one loop so that divergences remain; it is only at
higher
loops that this method regulates [9]. Higher derivatives must therefore be used
in conjunction with some other technique to provide complete regularization.
One
example of such a hybrid scheme has been given by Slavnov [10].

Nonlocalization results in propagators which fall off exponentially in
Euclidean
momentum space, as opposed to the power law fall off of the higher derivative
method. A direct consequence is that the higher derivative propagator contains
new poles, some of which necessarily represent ghost particles. Nonlocalization
introduces no new singularities. This is why nonlocalization can preserve
perturbative unitarity {\it before the unregulated limit is taken} while the
higher derivative method cannot. Although both methods induce new interactions
those from nonlocalization are no worse behaved than the interactions of the
original theory; where they have more derivatives there are also compensating
inverse powers of the regularization scale. (This is why nonlocalized loop
amplitudes are consistent with power counting in the local theory.) Another
important difference between the two methods is that higher derivatives
preserve
local gauge invariance whereas nonlocalization preserves only its physical
import, namely decoupling. The actual transformation law is generally distorted
into a nonlocal and nonlinear rule.

The second scheme with which we should contrast nonlocalization is {\it
continuum regularization} [11]. In this method spacetime is endowed with an
extra dimension and the fields are determined classically in terms of
stochastic
fields, of which there is one for each of the old fields. One takes expectation
values of products of fields on the higher dimensional spacetime by simply
expanding the fields in terms of the stochastic variables and then substituting
the appropriate stochastic correlation functions. The regularization is imposed
by nonlocalizing the correlation functions. Regulated Green's functions in the
lower dimensional theory are obtained by taking the extra coordinate to
infinity.

Although the role of the stochastic correlation functions is somewhat similar
to
the smearing operator of nonlocalization the two methods have many differences.
Nonlocalization leaves the tree amplitudes of any theory unchanged whereas
continuum regularization does not. Nonlocalization distorts gauge invariance
into a nonlocal and nonlinear symmetry which only closes on shell whereas
continuum regularization preserves the original transformation rule.
Nonlocalization produces an action in the original dimension whereas continuum
regularization is not derivable from an action on this spacetime. Finally, if
a suitable measure factor exists then nonlocalization gives a theory which is
perturbatively unitary {\it before the unregulated limit is taken} whereas this
is not the case for continuum regularization. We wish to emphasize that our
point in making this comparison is that the two methods differ, not that one is
superior to the other. In fact it seems to us that which method is preferable
depends very much on what one desires to do; and it is in any case useful to
have different perspectives from which the same problem can be viewed.

\vskip 1cm
\centerline{\bf 2. General first stage construction}

We will describe our method in terms of a generic action $S$ which is a
functional of generalized fields $\phi_i$. These fields may be commuting or
anticommuting, or there may be some of each type. The index which we write as
$i$ may represent any sort of label, including Lorentz tensor or spinor
indices.
The only restriction is that the theory must make sense perturbatively, that
is,
the action can be decomposed into free and interacting parts:
$$S[\phi] = F[\phi] + I[\phi] \eqno(2.1a)$$
$$F[\phi] =\half\int d^Dx \enskip \phi_i\ {\cal F}_{ij}\ \phi_j \eqno(2.1b)$$
where $I[\phi]$ is analytic in $\phi$ around the point $\phi = 0$.

\def\cale#1{{\cal E}_{#1}}
\def\calo#1{{\cal O}_{#1}}
\def\esq{{\cal E}^2}

Our construction for nonlocalizing $S[\phi]$ depends upon the specification of
a smearing operator $\cale{ij}$. For simplicity we shall take this to have the
simple exponential form:
$$\cale{}\equiv\exp\left[{{\cal F}\over 2\Lambda^2}\right]\eqno(2.2)$$
but the construction would work as well for any entire function of ${\cal F}
\Lambda^{-2}$ which is unity for ${\cal F} = 0$ and which falls faster than any
power of ${\cal F}$. We settled upon (2.2) only because it makes loop integrals
simple. The regularization parameter is $\Lambda$; taking $\Lambda \rightarrow
\infty$ gives the unregulated limit.

Two useful quantities which follow from the smearing operator are:
$${\widehat \phi}_i \equiv \cale{ij}^{-1} \phi_j \eqno(2.3)$$
$$\calo{} \equiv \left(\esq - 1\right) {\cal F}^{-1} = \int_0^1 {d\tau \over
\Lambda^2}\ {\cal E}^{2\tau}\eqno(2.4)$$
Note that $\calo{} $ is always well defined and invertible, even for ${\cal F}
=
0$. It is especially significant that $\calo{}$ still makes sense when
${\cal F}$ is the kinetic operator of a gauge theory.

For each field $\phi_i$ we now introduce an auxiliary field $\psi_i$ of the
same type. To avoid confusion with other auxiliaries which may be present we
shall call the $\psi$'s, ``shadow fields.'' The auxiliary action which couples
$\phi_i$ and $\psi_i$ is:
$${\cal S}[\phi,\psi] \equiv F[{\widehat \phi}~] - A[\psi] + I[\phi + \psi]
\eqno(2.5a)$$
$$A[\psi] \equiv \half \int d^Dx \enskip \psi_i \thinspace \calo{ij} \thinspace
\psi_j \eqno(2.5b)$$
We prove in appendix A that the classical shadow field equations:
$${\delta {\cal S}[\phi,\psi] \over \delta \psi_i(x)} = 0 \eqno(2.6a)$$
uniquely determine the shadow fields as functionals, $\psi_i[\phi]$, of the
original fields $\phi_i$ --- up to terms which vanish with the $\phi_i$ field
equations. Multiplying by the invertible operator $\calo{}$:
$$\psi_i = \calo{ij} \thinspace {\delta I[\phi + \psi] \over \delta \psi_j}
\eqno(2.6b)$$
gives an expression from which we can develop a series expansion for
$\psi_i[\phi]$. In fact the quantity $(\phi + \psi)$ is nothing more than the
classical field for the action $-A + I$ in the presence of the background
$\phi$.

The nonlocalized action is obtained by substituting this classical solution
into the auxiliary action:
$${\widehat S}[\phi] \equiv {\cal S}\Bigl[\phi,\psi[\phi]\Bigr] \eqno(2.7)$$
Expanding ${\widehat S}$ in powers of $\phi$ gives the kinetic term
$F[{\widehat \phi}~]$ plus an infinite series of interaction terms, the first
of which is just $I[\phi]$. There is a simple graphical representation for the
higher interaction vertices: they are the connected trees which follow from
using the local interaction vertex but replacing the propagator by $-i
\calo{}$.
Note that since $\calo{}$ is an entire function of ${\cal F}$ so too are these
higher interactions. This point is crucial in seeing that the method can
preserve perturbative unitarity.

We show in appendix B that if the infinitesimal transformation:
$$\delta \phi_i = T_i[\phi] \eqno(2.8a)$$
generates a symmetry of $S[\phi]$ then the following transformation generates a
symmetry of ${\widehat S}[\phi]$:
$${\widehat \delta} \phi_i = {\cal E}^2_{ij} \thinspace T_j\Bigl[\phi +
\psi[\phi]\Bigr] \eqno(2.8b)$$
Therefore nonlocalization preserves generally distorted versions of any and all
continuous symmetries. Another important result derived in appendix B is that
the shadow field transforms as follows:
$${\widehat \delta} \psi_i[\phi] = \Bigl(1 - {\cal E}^2\Bigr)_{ij} \thinspace
T_j\Bigl[\phi + \psi[\phi]\Bigr] - K_{ij}\Bigl[\phi + \psi[\phi]\Bigr]
\thinspace {\delta T_k \over \delta \phi_j}\Bigl[\phi + \psi[\phi]\Bigr]
\thinspace {\cal E}^2_{k\ell} \thinspace {\delta {\widehat S}[\phi] \over
\delta \phi_{\ell}} \eqno(2.9a)$$
$$K^{-1}_{ij}[\phi] \equiv {\cal O}^{-1}_{ij} - {\delta^2 I[\phi] \over \delta
\phi_i \delta \phi_j} \eqno(2.9b)$$
A direct consequence of the second term in (2.9a) is that closure of the local
symmetry --- by which we mean $[\delta_1,\delta_2] = \delta_3$ --- only implies
on shell closure:
$$[{\widehat \delta}_1,{\widehat \delta}_2] \thinspace \phi_i = {\widehat
\delta}_3 \thinspace \phi_i + {\cal E}^2_{ij} \thinspace \Omega^{12}_{jk}\Bigl[
\phi + \psi[\phi]\Bigr] \thinspace {\cal E}^2_{k\ell} \thinspace {\delta
{\widehat S}[\phi] \over \delta \phi_{\ell}} \eqno(2.10a)$$
$$\Omega^{12}_{i\ell}[\phi] \equiv {\delta T^1_i[\phi] \over \delta \phi_j}
\thinspace K_{jk}[\phi] \thinspace {\delta T^2_{\ell}[\phi] \over \delta
\phi_k}
- {\delta T^2_i[\phi] \over \delta \phi_j} \thinspace K_{jk}[\phi] \thinspace
{\delta T^1_{\ell}[\phi] \over \delta \phi_k} \eqno(2.10b)$$
It follows that enforcing off shell closure generally requires the inclusion of
generators which vanish with the field equations. In this sense the
nonlocalized
symmetry algebra is not only distorted but also somewhat enlarged.

When a field of the local theory does not appear in the transformation
functional $T_i[\phi]$ there is no reason it need be endowed with a shadow. The
nonlocalization of these fields can be accomplished merely by smearing the
kinetic term. An example is provided by the vector potential of QED. The great
advantage of this shortcut in QED is that it allows one to solve explicitly for
the shadow electron field. This is why a closed form for the action of nonlocal
QED could be given [7]. One disadvantage of the shortcut is that the associated
nonlocalized field equations possess nonperturbative, higher derivative
solutions of the same sort that wreck string theory [8].

Quantization is accomplished through the definition:
$$\Bigl\langle T^*\Bigl(O[\phi]\Bigr)\Bigr\rangle_{\cal E} \equiv \fint [D\phi]
\thinspace \mu[\phi] \thinspace \Bigl({\rm Gauge\ fixing}\Bigr) \thinspace
O[{\widehat \phi}~] \thinspace \exp\Bigl(i \thinspace {\widehat S}[\phi]\Bigr)
\eqno(2.11)$$
The object on the left is the nonlocally regulated vacuum expectation value
of the ${\rm T}^*$-ordered product of an arbitrary operator $O[\phi]$. It has a
subscript ${\cal E}$ to signify the dependence of nonlocalization upon the
choice of smearing operator. The two undefined quantities on the right are the
measure factor, $\mu[\phi]$, and the gauge fixing corrections. Both of these
are needed to maintain perturbative unitarity in gauge theories; for scalars
with nonderivative interactions there is no gauge symmetry to fix and the
measure factor can be set to unity.\footnote{*}{Experience with the nonlinear
sigma model leads us to expect that scalars with derivative interactions
require
a nontrivial measure factor in order to maintain perturbative unitarity.}

The issue of perturbative unitarity was discussed extensively in the treatment
of QED [7]. The basic idea is that the Cutkosky rules apply to nonlocal
theories whose propagators have poles only where ${\cal F} = 0$ and whose
interaction vertices are entire functions of ${\cal F}$. Since ${\widehat S}[
\phi]$ meets both requirements this guarantees unitarity on a large space of
states which includes unphysical polarizations. If the quantum theory is gauge
invariant then these unphysical polarizations decouple and it is also unitary
on
the physical space of states. We will explain in the next section that it is
simplest to gauge fix the local action and then nonlocalize the resulting BRS
theory. Since our procedure makes no distinction between invariant actions and
BRS gauge-fixed ones it follows that a nonlocalized version of BRS invariance
is induced.

If the local action is invariant under a continuous symmetry $\delta$ we have
seen that the nonlocalized action will be invariant under the symmetry
${\widehat \delta}$ which is given in equation (2.8b). The quantum theory will
possess this symmetry if and only if a suitable measure factor can be found
such that ${\widehat \delta} \Bigl([D \phi] \thinspace \mu[\phi]\Bigr) = 0$.
This condition can be reexpressed as follows using simple matrix manipulations
and relation (B.12b):
$$\eqalignno{{\widehat \delta} \Biggl\{ \ln\Bigl(\mu[\phi]\Bigr) \Biggr\}&=
- {\rm Tr}\Biggl\{{\delta {\widehat \delta} \phi_i \over \delta \phi_m}\Biggr\}
&(2.12a) \cr
&= -{\rm Tr}\Biggl\{{\cal E}^2_{ij} \thinspace {\delta T_j \over \delta \phi_k}
\Bigl[\phi + \psi[\phi]\Bigr] \thinspace K_{k\ell}\Bigl[\phi + \psi[\phi]\Bigr]
\thinspace {\cal O}^{-1}_{\ell m}\Biggr\} &(2.12b) \cr}$$
In addition the interaction vertices of the measure factor must be entire
functions of the operator ${\cal F}_{ij}$, and of course they must not spoil
the
regularization. Such a choice may not exist. In this case one of the local
symmetries is potentially anomalous; one must renormalize and take the
unregulated limit to be sure. An explicit example of the appearance of such an
anomaly in nonlocal regularization has been described by Hand in the chiral
Schwinger model [12].

A curiosity of definition (2.11) is that the field functions on its right hand
side are hatted while the field operators on the left are not. The reason for
this can be seen by taking $O[\phi]$ to be a simple product:
$$O[\phi] \equiv \phi_{i_1} \thinspace \phi_{i_2} \thinspace \dots \thinspace
\phi_{i_n} \eqno(2.13)$$
and then acting ${\cal F}_{j i_1}$ on both sides of (2.11). The zeroth order
result on the left is just the various commutator terms; the right hand side
gives a sum over pairings of propagators, plus interaction terms.
Correspondence
at zeroth order implies that a factor of ${\cal E}^{-1}$ resides on each field
in the functional integrand.

We emphasize that it is important to eliminate the shadow fields at the
classical level. Functionally integrating over them would engender divergences
from shadow loops; nor does perturbative unitarity require us to quantize these
fields because their propagators contain no poles. Nevertheless, the Feynman
rules of ${\widehat S}[\phi]$ are very cumbersome because of all the induced
interactions while the Feynman rules of ${\cal S}[\phi,\psi]$ are essentially
as simple as those of the local theory. We have therefore chosen to use the
simple Feynman rules of ${\cal S}[\phi,\psi]$ and to enforce the condition that
$\psi_i$ obey its classical field equation ($\psi_i = \psi_i[\phi]$) by merely
omitting closed loops composed solely of shadow lines.

The resulting nonlocalized Feynman rules are a trivial extension of the local
ones. Except for the measure factor the vertices are unchanged, but every leg
can now connect either to a smeared propagator:\footnote{*}{A digression is
necessary to comment on the various factors of $i$ which appear in what the
reader was assured would be a Euclidean formulation of perturbation theory.
{}From
the earliest work on nonlocal regularization [7] it has been convenient to set
perturbation theory up in Minkowski space and define loop integrations by
formal
Wick rotation. This convention will be followed here as well. We stress that it
is only a trick to facilitate the analytic continuation. If rigor is desired
then the whole thing can be done in Euclidean space where ${\cal F}$ is a
nonpositive operator.}
$${i {\cal E}^2\over {\cal F} + i\epsilon} = -i \int_1^{\infty} {d\tau\over
\Lambda^2} \enskip \exp\Bigl(\tau {{\cal F}\over \Lambda^2}\Bigr)
\eqno(2.14a)$$
or to a shadow propagator:
$$-i {\cal O} = {i(1-{\cal E}^2)\over {\cal F}} = -i\int_0^1 {d\tau \over
\Lambda^2} \enskip \exp\Bigl(\tau {{\cal F}\over \Lambda^2}\Bigr)
\eqno(2.14b)$$
The two sorts of propagators are represented graphically for Yang-Mills theory
in figures 1a and 1b. We place a bar across the shadow line because its
propagator lacks a pole and so carries no quanta. For this reason all external
lines must be unbarred. (This more than compensates the factors of ${\cal E}^{
-1}$ which arise because we use $O[{\widehat \phi}~]$ rather than $O[\phi]$ on
the right hand side of equation (2.11).) Vertices from the measure factor are
represented graphically with a circled ``M,'' as in figure 2a. Measure vertices
can be connected only to barred lines. The symmetry factor of any diagram is
computed without distinguishing between barred and unbarred lines.

A number of points deserve mention. First, tree order Green's functions are
unchanged except for external line factors which are unity on shell. This
follows because every internal line of a tree graph can be either barred or
unbarred. Hence it is only the sum of (2.14a) and (2.14b) which enters, and
this
sum just gives the local propagator. Second, the fact that on-shell tree
amplitudes are unchanged by nonlocalization means that ${\widehat S}[\phi]$
preserves {\it any and all} symmetries of $S[\phi]$ on shell. A direct proof of
this fact is given for continuous symmetries in appendix B. Third, all loops
contain at least one of the smeared propagators (2.14a) and are therefore
convergent in the ultraviolet. Finally, when applying the cutting rules one
allows only unbarred lines to be cut because they alone contain poles. This
automatically implies unitarity on the extended space of states which includes
unphysical polarizations.

When the measure factor is unity nonlocal regularization is no more difficult
to
use than dimensional regularization. We have shown this by carrying out the
full
one and two loop renormalization program for scalar $\phi^4$ theory in four
spacetime dimensions and for $\phi^3$ theory in six dimensions [13]. We refer
the interested reader to this work for a very detailed study of the method in
operation. In particular it should be noted that renormalization proceeds in
nonlocal regularization as in any other method: the bare parameters are
determined as functions of the renormalized ones and the scale $\Lambda$ such
that the limit $\Lambda \longrightarrow \infty$ is finite for all noncoincident
Green's functions. Even though ${\widehat S}[\phi]$ possesses nonlocal
interactions its bare parameters are still those of a local theory. Once a
smearing operator has been chosen the nonlocal interactions are completely
determined by the local ones. We remark also that the method correctly handles
overlapping divergences and that power counting works as it would if the scale
$\Lambda$ had been a momentum cutoff.

\vskip 1cm
\centerline{\bf 3. The second stage for Yang-Mills}

When vector and/or tensor quanta are allowed to interact some form of gauge
invariance must be present in order to reconcile Poincar\'e invariance and
perturbative unitarity. For this invariance to continue operating in loops the
functional formalism must respect it. This generally requires that the
functional measure factor should depend nontrivially upon the fields. The
measure factor must also meet the requirements for being an acceptable
interaction in the regulated theory. This means that it must have manifest
Poincar\'e invariance, that it can involve only entire functions of the
kinetic operator, and that diagrams involving it must be well damped in
Euclidean momentum space. If no such measure factor exists then the regulated
theory has an anomaly.\footnote{*}{The anomaly may, however, vanish in the
unregulated limit.} For the appearance of such an anomaly in nonlocal
regularization we refer the reader to Hand's study of the chiral Schwinger
model [12].

We shall take pure Yang-Mills for the unregulated theory. Its Lagrangian and
field strength tensor are:
$${\cal L} = -\frac14 \thinspace F_{a\mu\nu} \thinspace F^{\mu\nu}_a
\eqno(3.1a)$$
$$F_{a\mu\nu} \equiv A_{a\nu,\mu} - A_{a\mu,\nu} - g f_{abc} \thinspace
A_{b\mu} \thinspace A_{c\nu} \eqno(3.1b)$$
where $g$ is the coupling constant and the $f_{abc}$ are the structure
constants. It is invariant under the familiar transformation:
$$\delta_{\theta} A_{a\mu} = - \theta_{a,\mu} + g f_{abc} \thinspace A_{b\mu}
\thinspace \theta_c \eqno(3.2)$$
Note that the local symmetry algebra closes for all field configurations:
$$[\delta_{\theta_1},\delta_{\theta_2}] \thinspace A_{a\mu} = -\Theta_{a,\mu}
+ g f_{abc} \thinspace A_{b\mu} \thinspace \Theta_{c} \eqno(3.3a)$$
$$\Theta_a \equiv g f_{abc} \thinspace \theta_{1b} \thinspace \theta_{2c}
\eqno(3.3b)$$

To nonlocalize we identify the kinetic operator:
$${\cal F}^{\alpha \beta}_{ab} \equiv \delta_{ab} \thinspace (\partial^2
\eta^{\alpha \beta} - \partial^{\alpha} \partial^{\beta}) \eqno(3.4a)$$
and the interaction:
$$I[A] \equiv \int d^Dx \thinspace \Biggl\{g f_{abc} \thinspace A_{a\nu,\mu}
\thinspace A_b^{~\mu} \thinspace A_c^{~\nu} - \frac14 g^2 f_{abc} f_{cde}
\thinspace A_{a\mu} \thinspace A_{b\nu} \thinspace A_d^{~\mu} \thinspace
A_e^{~\nu} \Biggr\} \eqno(3.4b)$$
Comparison with (2.5) and (2.6) gives the following expansion for the shadow
field:
$$\eqalign{B_a^{~\alpha}[A] &= {\cal O}_{ab}^{\alpha \beta} \thinspace {\delta
I[A + B] \over \delta A_b^{~\beta}} \cr
&= {\cal O}_{ab}^{\alpha \beta} \thinspace g f_{bcd} \thinspace \Bigl[
A_c^{~\beta} \thinspace A^{~\gamma}_{d~,\gamma} + A_{c \gamma} \thinspace
A_d^{~\gamma,\beta} - 2 \thinspace A_{c \gamma} \thinspace A_d^{~\beta,\gamma}
\Bigr] + O(g^2 A^3) \cr} \eqno(3.5)$$
The resulting nonlocalized action follows immediately from (2.5) and (2.7):
$${\widehat S}[A] = \half \int d^Dx \Biggl\{ {\widehat A}_{a \alpha} \thinspace
{\cal F}^{\alpha \beta}_{ab} \thinspace {\widehat A}_{b \beta}
- B_{a \alpha}[A] \thinspace {\cal O}^{-1~\alpha \beta}_{~~~ ab} \thinspace
B_{b \beta}[A] \Biggr\} + I\Bigl[A + B[A]\Bigr] \eqno(3.6)$$

We see from (2.8) that the nonlocalized gauge symmetry is:
$$\eqalignno{{\widehat \delta}_{\theta} A_a^{~\alpha} &= {\cal E}^{2~ \alpha
\beta}_{~~~ ab} \thinspace \Biggl\{-\theta_a^{~,\alpha} + g f_{bcd} \thinspace
\Bigl(A_{c \beta} + B_{c \beta}[A]\Bigr) \thinspace \theta_d\Biggr\} \cr
&=- \theta_a^{~,\alpha} + {\cal E}^{2~ \alpha \beta}_{~~~ ab} \thinspace g
f_{bcd} \thinspace \Bigl(A_{c \beta} + B_{c \beta}[A]\Bigr) \thinspace \theta_d
&(3.7) \cr}$$
As was the case for QED [7], nonlocalizing Yang-Mills results in a gauge
transformation which is neither linear nor local. Another point in common with
QED is the failure of the symmetry generators to close under commutation,
except
for field configurations which obey the equations of motion. Comparison with
relations (2.9) and (2.10) shows:
$$[{\widehat \delta}_{\theta_1},{\widehat \delta}_{\theta_2}]
\thinspace A_a^{~\alpha} = {\widehat \delta}_{\Theta} A_a^{~\alpha}
+ {\cal E}^{2~ \alpha \beta}_{~~~ ab} \thinspace \Omega_{bc \beta \gamma}
\Bigl[A + B[A]\Bigr] \thinspace {\cal E}^{2~ \gamma \delta}_{~~~ cd} \thinspace
{\delta {\widehat S}[A] \over \delta A_d^{~\delta}} \eqno(3.8a)$$
$$\Omega^{\alpha \beta}_{ab}[A] \equiv g^2 \thinspace f_{acd} \thinspace
f_{bef}
\thinspace \Biggl\{\theta_{1d} \thinspace K^{\alpha \beta}_{ce}[A] \thinspace
\theta_{2f} - \theta_{2d} \thinspace K^{\alpha \beta}_{ce}[A] \thinspace
\theta_{1f}\Biggr\} \eqno(3.8b)$$
$$\Bigl(K^{\alpha \beta}_{ab}\Bigr)^{-1}[A] = {\cal O}^{-1
{}~\alpha\beta}_{~~~ab}
- {\delta^2 I[A] \over \delta A_{a \alpha} \delta A_{b \beta}} \eqno(3.8c)$$
This means that the space of generators (3.7) is only a subset of a true
algebra which includes generators that vanish with the field equations.

The most obvious application of the procedure laid out in section 2 would be to
functionally quantize as follows:
$$\Bigl\langle T^*\Bigl(O[A]\Bigr)\Bigr\rangle_{\cal E} = \fint [DA] \thinspace
\mu[A] \thinspace \Bigl({\rm Gauge\ Fixing}\Bigr) \thinspace O[{\widehat A}~]
\thinspace \exp\Bigl(i\thinspace {\widehat S}[A]\Bigr) \eqno(3.9)$$
The trivial problem with this approach is that the measure factor defined by
equation (2.12) would be singular on account of the unsuppressed, longitudinal
components of transformation law (3.7). This could be circumvented by the
simple
expedient of redefining what we called the ``gauge parameter'' so as to absorb
the unsuppressed terms into the zeroth order piece.

The harder problem is how to fix the gauge. It is here that the nonclosure of
${\widehat \delta}$ has its only significant impact. We might parallel the
usual
derivation of Faddeev and Popov by introducing unity in the form:
$$1 = \fint [D\theta] \enskip \delta\Bigl[\partial \cdot A_a^{\theta} - f_a
\Bigr] \thinspace \det\Biggl({\delta \partial \cdot A_b^{\theta} \over \delta
\theta_c}\Biggr) \eqno(3.10)$$
where by $A^{\theta}$ we mean the transformed vector potential. One would then
change variables $A \rightarrow A'=A^{\theta}$ and exploit the gauge invariance
of ${\widehat S}[A]$ and $[DA] \thinspace\mu[A]$, {\it and the presumed
invariance of the determinant}, to factor out the functional integration over
$\theta$. The final step would be to functionally average over $f_a$ with the
desired weighting factor, usually a simple Gaussian. Unfortunately the
Faddeev-Popov determinant is not invariant owing to the failure of two
$\delta$-transformations to give another $\delta$-transformation. Hence the
integration over $\theta$ does not quite factor out and the naive derivation
fails.

This problem is a familiar one from supergravity and the answer is well known
[14]. The physically essential symmetry of a gauge-fixed theory is BRS
invariance. When the algebra of gauge transformations closes for all field
configurations the the ghost structure of the BRS action comes from
exponentiating the Faddeev-Popov determinant, and the BRS transformation of the
gauge field is just a gauge transformation with the parameter equal to a
constant Grassmann number times the ghost field. For an algebra which fails to
close off shell one needs to introduce higher ghost terms into both the action
and the BRS transformation. There will also generally need to be a correction
to the measure factor. One could probably derive the necessary changes by
formally integrating the Faddeev-Popov determinant with respect to $\theta$,
however a much simpler approach is to nonlocalize the BRS Lagrangian directly.

In Feynman gauge the local BRS Lagrangian is:
$$\eqalign{{\cal L}_{\rm\scriptscriptstyle BRS} =-\half A_{a\nu,\mu} \thinspace
&A_a^{~\nu,\mu} - {\overline \eta}_a^{~,\mu} \thinspace \eta_{a,\mu} + g
\thinspace f_{abc}\thinspace {\overline \eta}_a^{~,\mu} \thinspace A_{b\mu}
\thinspace \eta_c \cr
& + g f_{abc} \thinspace A_{a\nu,\mu} \thinspace A_b^{~\mu}
\thinspace A_c^{~\nu} - \frac14 g^2 f_{abc} f_{cde} \thinspace A_{a\mu}
\thinspace A_{b\nu} \thinspace A_d^{~\mu} \thinspace
A_e^{~\nu}\cr}\eqno(3.11)$$
It is invariant under the following global symmetry transformation:
$$\delta A_{a \alpha} = \Bigl(\eta_{a,\alpha} - g f_{abc} \thinspace
A_{b\alpha}
\thinspace \eta_c\Bigr) \thinspace \delta \zeta \eqno(3.12a)$$
$$\delta \eta_a = - \half g \thinspace f_{abc} \thinspace \eta_b \thinspace
\eta_c \thinspace \delta \zeta \eqno(3.12b)$$
$$\delta {\overline \eta}_a = - A^{~\mu}_{a~,\mu} \thinspace \delta \zeta
\eqno(3.12c)$$
where $\zeta$ is a constant anticommuting $\comp$-number.

To nonlocalize this action we identify the gluon and ghost kinetic operators:
$${\cal F}^{\alpha \beta}_{ab} = \delta_{ab} \thinspace \eta^{\alpha \beta}
\thinspace \partial^2 \eqno(3.13a)$$
$${\cal F}_{ab} = \delta_{ab} \thinspace \partial^2 \eqno(3.13b)$$
Since the indices and derivatives separate we have found it convenient to use
a smearing operator with no indices:
$${\overline {\cal E}} \equiv \exp\Bigl({\partial^2 \over 2 \Lambda^2}\Bigr)
\eqno(3.14a)$$
$${\overline {\cal O}} \equiv {{\overline {\cal E}}^2 - 1 \over \partial^2}
\eqno(3.14b)$$
The BRS interaction is:
$$I[A,{\overline \eta},\eta] = g \thinspace f_{abc} \int d^Dx \thinspace
\Biggl\{ {\overline \eta}_a^{~,\mu} \thinspace A_{b \mu} \thinspace \eta_c
+ A_{a\nu,\mu} \thinspace A_b^{~\mu} \thinspace A_c^{~\nu} - \frac14 g f_{cde}
\thinspace A_{a\mu} \thinspace A_{b\nu} \thinspace A_d^{~\mu} \thinspace
A_e^{~\nu}\Biggr\} \eqno(3.15)$$
Comparison with (2.5) and (2.6) gives the following expansions for the shadow
fields:
$$\eqalign{B_a^{~\alpha}[A,{\overline \eta},\eta] &= {\overline {\cal O}}
\thinspace {\delta I \over \delta A_{a\alpha}}\Bigl[A + B, {\overline \eta}
+ {\overline \psi},\eta + \psi\Bigr] \cr
&= {\overline {\cal O}} \thinspace g f_{abc} \thinspace \Bigl[A_b^{~\alpha}
\thinspace A^{~\beta}_{c,~\beta} + A_{b \beta} \thinspace A_c^{~\beta,\alpha}
- 2 \thinspace A_{b \beta} \thinspace A_c^{~\alpha,\beta} - {\overline
\eta}_b^{
{}~,\alpha} \thinspace \eta_c\Bigr] + O(g^2) \cr} \eqno(3.16a)$$
$$\eqalign{{\overline \psi}_a[A,{\overline \eta},\eta] &= - {\overline {\cal
O}}
\thinspace {\delta I \over \delta \eta_a}\Bigl[A + B, {\overline \eta} +
{\overline \psi},\eta + \psi\Bigr] \cr
&= -{\overline {\cal O}} \thinspace g f_{abc} \thinspace A_{b\alpha} \thinspace
{\overline \eta}_c^{~\alpha} + O(g^2) \cr} \eqno(3.16b)$$
$$\eqalign{\psi_a[A,{\overline \eta},\eta] &= {\overline {\cal O}} \thinspace
{\delta I \over \delta {\overline \eta}_a}\Bigl[A + B, {\overline \eta} +
{\overline \psi},\eta + \psi\Bigr] \cr
&=-{\overline {\cal O}} \thinspace g f_{abc} \thinspace \partial^{\alpha}
\Bigl(
A_{b\alpha} \thinspace \eta_c\Bigr) + O(g^2) \cr} \eqno(3.16c)$$
To economize space we will henceforth suppress the functional argument list on
the shadow fields. From expressions (2.5) and (2.7) we see that the
nonlocalized
BRS action is:
$$\eqalign{{\widehat S}[A,{\overline \eta},\eta] = \int d^Dx \thinspace
\Biggl\{
-\half {\widehat A}_{a \alpha, \beta} \thinspace {\widehat A}_a^{~\alpha,
\beta}
- \half B_{a \alpha} \thinspace &{\overline {\cal O}}^{~-1} \thinspace B_a^{~
\alpha} - {\widehat {\overline \eta}}_a^{~,\alpha} \thinspace {\widehat
\eta}_{a
,\alpha} - {\overline \psi}_a \thinspace {\overline {\cal O}}^{~-1} \thinspace
\psi_a\Biggr\} \cr
&+ I\Bigl[A+B, {\overline \eta}  + {\overline \psi}, \eta + \psi\Bigr] \cr}
\eqno(3.17)$$

The nonlocalized BRS symmetry which follows from applying (2.8) to (3.12) is
easy to give in terms of the shadow fields:
$${\widehat \delta} A_{a \alpha} = {\overline {\cal E}}^2 \thinspace \Bigl\{
(\eta_{a,\alpha} + \psi_{a,\alpha}) - g \thinspace f_{abc} \thinspace
(A_{b \alpha} + B_{c \alpha}) \thinspace (\eta_c + \psi_c)\Bigr\} \thinspace
\delta \zeta \eqno(3.18a)$$
$${\widehat \delta} \eta_a = -\half g \thinspace f_{abc} \thinspace {\overline
{\cal E}}^2 \thinspace (\eta_b + \psi_b) \thinspace (\eta_c + \psi_c)
\thinspace
\delta \zeta \eqno(3.18b)$$
$${\widehat \delta} {\overline \eta}_a = - {\overline {\cal E}}^2 \thinspace
(A_{a~,\alpha}^{~\alpha} + B_{a~,\alpha}^{~\alpha}) \thinspace \delta \zeta
\eqno(3.18c)$$
Because the shadow fields --- $B_{a \alpha}[A,{\overline \eta},\eta]$,
${\overline \psi}_a[A,{\overline \eta},\eta]$ and $\psi_a[A,{\overline \eta},
\eta]$ --- depend upon the ghosts not every occurrence of $\eta_a$ in (3.18a)
derives from the replacement $\theta_a \longrightarrow -\eta_a \thinspace
\delta \zeta$ in (3.7). These ``extra'' terms appear as well in the BRS
transformations of supergravity models whose algebras only close on shell [14].

The transformation law (3.18) is inconvenient because the variation of the
antighost no longer produces just a longitudinal gluon. This property plays a
special role in using BRS invariance to prove decoupling. We therefore fold in
a
dynamically irrelevant symmetry of the class described in equations (B.16) and
(B.17):
$${\widehat \delta}^{~0} A_{a \alpha} = (1 - {\overline {\cal E}}^2) \thinspace
\eta_{a,\mu} \delta \zeta - {\overline {\cal E}}^2 \thinspace \psi_{a,\mu}
\thinspace \delta \zeta \eqno(3.19a)$$
$${\widehat \delta}^{~0} \eta_a = 0 \eqno(3.19b)$$
$${\widehat \delta}^{~0} {\overline \eta}_a = - (1- {\overline {\cal E}}^2)
\thinspace A_{a~,\mu}^{~\mu} \delta \zeta + {\overline {\cal E}}^2 \thinspace
B_{a~,\mu}^{~\mu} \thinspace \delta \zeta \eqno(3.19c)$$
Combining the two gives a physically equivalent but more convenient
transformation:
$${\widehat \delta}^{~1} A_{a \alpha} = \Biggl\{\eta_{a,\alpha} - g \thinspace
f_{abc} \thinspace {\overline {\cal E}}^2 \thinspace (A_{b \alpha}
+ B_{c \alpha}) \thinspace (\eta_c + \psi_c)\Biggr\} \thinspace \delta \zeta
\eqno(3.20a)$$
$${\widehat \delta}^{~1} \eta_a = -\half g \thinspace f_{abc} \thinspace
{\overline {\cal E}}^2 \thinspace (\eta_b + \psi_b) \thinspace (\eta_c +
\psi_c)
\thinspace \delta \zeta \eqno(3.20b)$$
$${\widehat \delta}^{~1} {\overline \eta}_a = - A_{a~,\mu}^{~\mu} \thinspace
\delta \zeta \eqno(3.20c)$$
That such a form can be reached also follows from the general argument of de
Wit and van Holten [14].

Since the formalism of section 2 applies to gauge fixed theories as well as to
invariant ones we can read off the rule for functional quantization directly
from expression (2.11):
$$\Bigl\langle T^*\Bigl(O[A,{\overline \eta},\eta]\Bigr)\Bigr\rangle_{\cal E}
\equiv \fint [DA] [D{\overline \eta}] [D\eta] \enskip \mu[A,{\overline \eta},
\eta] \thinspace O[{\widehat A},{\widehat {\overline \eta}},{\widehat \eta}]
\thinspace \exp\Bigl(i\thinspace {\widehat S}[A,{\overline \eta},\eta] \Bigr)
\eqno(3.21)$$
Note, however, that it should be the transformation ${\widehat \delta}^{~1}$,
not ${\widehat \delta}$, which goes into expression (2.12) to partially
determine the measure factor. Of course one can freely modify $\mu[A,{\overline
\eta},\eta]$ by terms that are invariant under ${\widehat \delta}^{~1}$ but
with
a natural sort of minimality condition the answer is:
$$\ln\Bigl(\mu[A,{\overline \eta},\eta]\Bigr) = -\frac{i}2 g^2 \thinspace
f_{acd} f_{bcd} \int d^Dx \enskip A_{a\mu} \thinspace {\cal M} \thinspace
A^{\mu}_b + O(g^3) \eqno(3.22a)$$
$$\eqalignno{{\cal M} \equiv {1\over 2^D \thinspace \pi^{D/2}} \int_0^1 d{\tau}
\enskip &{\Lambda^{D-2}\over (\tau + 1)^{D/2}} \thinspace \exp\Bigl({\tau \over
 \tau + 1} \thinspace {\partial^2 \over \Lambda^2}\Bigr)\cr
&\times \Bigl\{{2\over \tau +1} - (D-1) + 2(D-1)\thinspace {\tau \over \tau+1}
\Bigr\} &(3.22b) \cr}$$
There is no proof that the requisite higher order terms exist, although this is
what we expect because the local theory is free of anomalies.

The presumed existence of a suitable measure factor implies Slavnov-Taylor
identities in the usual way. Consider, for example, the following choice for
the operator $O[A,{\overline \eta},\eta]$ of (3.21):
$$O[A,{\overline \eta},\eta] = {\overline \eta}_a(x) \thinspace A^{~\nu}_{b~,
\nu}(y) \eqno(3.23)$$
Of course the functional integral vanishes by ghost number counting, but a BRS
transformation gives a nontrivial relation between Green's functions:
$${\widehat \delta}^{~1} \Biggl({\overline \eta}_a(x) \thinspace
A^{~\nu}_{b,\nu}(y)\Biggr) = - A^{\mu}_{a,\mu}(x) \thinspace A^{\nu}_{b,\nu}(y)
\thinspace \delta \zeta + {\overline \eta}_a(x) \thinspace {\delta {\widehat S}
\over \delta {\overline \eta}_b(y)} \thinspace \delta \zeta \eqno(3.24)$$
Upon functionally integrating by parts we infer the transversality of the
vacuum
polarization at one loop.\footnote{*}{Although we expect transversality to
persist to all orders we have no proof of this for lack of knowledge about the
higher order terms in the measure factor. In particular we have not been able
to
decide whether or not the measure depends upon the ghosts.} This result will be
explicitly verified in the next section.

\vskip 1cm
\centerline{\bf 4. The gluon self energy at one loop}

The Feynman rules can be read off from the functional formalism (3.21) using
relations (3.17) and (3.22). The barred and unbarred gluon propagators are
represented graphically in figure 1a and have the following forms:
$${-i \delta_{ab} \thinspace \eta^{\alpha \beta} \over p^2 - i \epsilon}
\thinspace \exp\left(-{p^2 \over \Lambda^2}\right) = -i \delta_{ab} \thinspace
\eta^{\alpha \beta} \thinspace \int_1^{\infty} {d\tau \over \Lambda^2} \enskip
\exp\left(-\tau {p^2 \over \Lambda^2}\right) \eqno(4.1a)$$
$${-i \delta_{ab} \thinspace \eta^{\alpha \beta} \over p^2 - i \epsilon}
\thinspace \left\{1 - \exp\left(-{p^2 \over \Lambda^2}\right) \right\} = -i
\delta_{ab} \thinspace \eta^{\alpha \beta} \thinspace \int_0^1 {d\tau \over
\Lambda^2} \enskip \exp\left(-\tau {p^2 \over \Lambda^2}\right) \eqno(4.1b)$$
The barred and unbarred ghost propagators are shown in figure 1b and can be
expressed as follows:
$${-i \delta_{ab} \over p^2 - i \epsilon} \thinspace \exp\left(-{p^2 \over
\Lambda^2}\right) = -i \delta_{ab} \thinspace \int_1^{\infty} {d\tau \over
\Lambda^2} \enskip \exp\left(-\tau {p^2 \over \Lambda^2}\right) \eqno(4.2a)$$
$${-i \delta_{ab} \over p^2 - i \epsilon} \thinspace \left\{1 - \exp\left(-{p^2
\over \Lambda^2}\right) \right\} = -i \delta_{ab} \thinspace \int_0^1 {d\tau
\over \Lambda^2} \enskip \exp\left(-\tau {p^2 \over \Lambda^2}\right)
\eqno(4.2b)$$
The obvious advantage of this gauge is that the tensor and gauge indices
decouple from the nonlocal factors. The 2-point measure vertex is represented
in figure 2a and has the form:
$$M^{\alpha\beta}_{ab}(p) \equiv - {f_{acd} \thinspace f_{bcd} \thinspace
\eta^{\alpha\beta}\over 2^D \thinspace (\pi)^{D/2}} \int_0^1 d\tau \enskip
{\Lambda^{D-2} \over (\tau + 1)^{D/2}} \thinspace \exp\Bigl(-{\tau \over
\tau+1}
\thinspace {p^2\over \lambda^2}\Bigr) \thinspace \Bigl[(D-1) -2{D-2 \over \tau
+1} \Bigr] \eqno(4.3)$$
The local 3-point vertices are shown in figure 2b and can be expressed as
follows:
$$I^{\alpha\beta\gamma}_{abc}(p_1,p_2,p_3) \equiv -i f_{abc} \Biggl\{
\eta^{\alpha \beta} \thinspace (p_2 - p_1)^{\gamma} + \eta^{\beta \gamma}
\thinspace (p_3 - p_2)^{\alpha} + \eta^{\gamma \alpha} \thinspace (p_1 - p_3)^{
\beta}\Biggr\} \eqno(4.4a)$$
$$I^{\alpha}_{abc}(p_1,p_2,p_3) \equiv i f_{abc} \thinspace p_3^{\alpha}
\eqno(4.4b)$$
The local 4-point vertex is depicted in figure 2c and has the form:
$$\eqalignno{I^{\alpha\beta\gamma\delta}_{abcd}(p_1,&p_2,p_3,p_4) \equiv
- \Biggl\{f_{abe} \thinspace f_{cde} \thinspace (\eta^{\alpha \gamma}
\thinspace
\eta^{\beta \delta} - \eta^{\gamma \beta} \thinspace \eta^{\delta \alpha}) \cr
&+ f_{ace} \thinspace f_{dbe} \thinspace (\eta^{\alpha \delta} \thinspace
\eta^{\gamma \beta} - \eta^{\delta \gamma} \thinspace \eta^{\alpha \beta})
+ f_{ade} \thinspace f_{bce} \thinspace (\eta^{\alpha \beta} \thinspace
\eta^{\delta \gamma} - \eta^{\beta \delta} \thinspace \eta^{\gamma \alpha})
\Biggr\} &(4.5) \cr}$$
The only thing missing is the higher measure vertices. Since the N-point
measure
vertex comes in at order $g^N$ it is clear that only for $N=2$ can it
contribute
to the (order $g^2$) one loop vacuum polarization.

The one loop vacuum polarization can be naturally divided into four parts.
These
correspond to figures 3-6 and we shall term them $A$, $B$, $C$ and $D$
respectively:
$$\Pi^{\alpha\beta}_{ab} = A^{\alpha\beta}_{ab} + B^{\alpha\beta}_{ab}
+ C^{\alpha\beta}_{ab} + D^{\alpha\beta}_{ab} \eqno(4.6)$$
The three barring variations depicted in figure 3 come from the 3-gluon vertex.
They can be summed up to give the following:
$$\eqalignno{i A^{\alpha\beta}_{ab}(p) = \half \int_{R^2_1} d^2\tau \int {d^Dk
\over (2\pi)^D} \enskip ig &I^{\alpha \gamma \delta}_{acd}(p,-k,-q) \thinspace
\Biggl\{-i \delta_{ce} \eta_{\gamma \rho} \exp\Bigl(- \tau_1 {k^2 \over
\Lambda^2}\Bigr) \Biggr\}\cr
& \times ig I^{\beta \rho \sigma}_{bef}(-p,k,q) \thinspace \Biggl\{-i
\delta_{df} \eta_{\delta \sigma} \exp\Bigl(- \tau_2 {q^2 \over \Lambda^2}\Bigr)
\Biggr\} &(4.7a) \cr}$$
$$\eqalignno{&= {i g^2 \thinspace f_{acd} \thinspace f_{bcd} \over 2^D
\thinspace\pi^{D/2}} \int_{R^2_1} {d^2\tau \thinspace \Lambda^{D-4} \over
(\tau_1 + \tau_2)^{D/2}} \thinspace \exp\Bigl(-{\tau_1 \thinspace \tau_2 \over
\tau_1 + \tau_2} \thinspace {p^2 \over \Lambda^2}\Bigr) \Biggl\{+\frac32 (D-1)
{\Lambda^2 \over \tau_1 + \tau_2} \eta^{\alpha \beta} \cr
& \hskip 3cm - {\tau_1 \thinspace \tau_2 \over \tau_1 + \tau_2} [p^2
\eta^{\alpha \beta} + (2D-3) p^{\alpha} p^{\beta}] + [\frac52 p^2 \eta^{\alpha
\beta} + \half (D-6) p^{\alpha} p^{\beta}]\Biggr\} &(4.7b) \cr}$$
The symbol $R^2_1$ in this expression denotes the standard 2-parameter region
of integration at one loop, namely the first quadrant minus the square $0 \leq
\tau_1, \tau_2 < 1$. A very similar contribution comes from the three ghost
loops of figure 4:
$$\eqalignno{i B^{\alpha\beta}_{ab}(p) = \int_{R^2_1} d^2\tau \int {d^Dk \over
(2\pi)^D} \enskip ig I^{\alpha}_{acd}(p&,-k,-q) \thinspace \Biggl\{-i
\delta_{ce} \exp\Bigl(- \tau_1 {k^2 \over \Lambda^2}\Bigr) \Biggr\}\cr
& \times ig I^{\beta}_{bef}(-p,k,q) \thinspace \Biggl\{-i \delta_{df}
\exp\Bigl(- \tau_2 {q^2 \over \Lambda^2}\Bigr) \Biggr\} &(4.8a) \cr}$$
$$={i g^2 \thinspace f_{acd} \thinspace f_{bcd} \over 2^D \thinspace \pi^{D/2}}
\int_{R^2_1} {d^2\tau \thinspace \Lambda^{D-4} \over (\tau_1 + \tau_2)^{D/2}}
\thinspace \exp\Bigl(-{\tau_1 \thinspace \tau_2 \over \tau_1 + \tau_2}
\thinspace {p^2 \over \Lambda^2}\Bigr) \thinspace \Biggl\{-\half {\Lambda^2
\over \tau_1 + \tau_2} \thinspace \eta^{\alpha \beta} + {\tau_1 \thinspace
\tau_2 \over \tau_1 + \tau_2} p^{\alpha} p^{\beta}\Biggr\} \eqno(4.8b)$$

The diagram of figure 5 comes from the 4-gluon vertex:
$$\eqalignno{i C^{\alpha\beta}_{ab}(p) &= \int {d^Dk \over (2\pi)^D} \enskip
ig^2 I^{\alpha \beta \gamma \delta}_{abcc}(p,-p,-k,k) \thinspace \Bigl({-i
\eta_{\gamma \delta} \over k^2 - i\epsilon}\Bigr) \exp\Bigl(-{k^2\over
\Lambda^2}\Bigr) &(4.9a) \cr
&=- {ig^2 \thinspace f_{acd} \thinspace f_{bcd} \over 2^D \thinspace \pi^{D/2}}
\thinspace \eta^{\alpha \beta} \thinspace 2\Bigl({D-1\over D-2}\Bigr)
\thinspace
\Lambda^{D-2} &(4.9b) \cr}$$
This contribution would vanish in dimensional regularization; here it plays an
essential role in canceling the quadratic divergence. The same comments pertain
as well to the measure factor contribution which comprises figure 6:
$$i D^{\alpha \beta}_{ab}(p) =- {ig^2 \thinspace f_{acd} \thinspace f_{bcd}
\over 2^D \thinspace \pi^{D/2}} \thinspace \eta^{\alpha \beta} \thinspace
\int_0^1 {d\tau \thinspace \Lambda^{D-2} \over (\tau + 1)^{D/2}} \thinspace
\exp\Bigl(-{\tau \over \tau + 1} \thinspace {p^2 \over \Lambda^2}\Bigr)
\Biggl\{(D-1) - 2 {D-2 \over \tau + 1}\Biggr\} \eqno(4.10)$$

To check that the vacuum polarization is transverse first add the longitudinal
parts of $A$ and $B$. Now reduce the parameter integral to which the sum is
proportional:
$$\eqalignno{\int_{R^2_1} {d^2\tau \over (\tau_1 + \tau_2)^{D/2}} \thinspace
&\exp\Bigl(-{\tau_1 \thinspace \tau_2 \over \tau_1 + \tau_2} \thinspace {p^2
\over \Lambda^2}\Bigr) \cr
&\times \Biggl\{(\frac32 D - 2) {1\over \tau_1 + \tau_2} - (2D - 3) {\tau_1
\thinspace \tau_2 \over (\tau_1 + \tau_2)^2} \thinspace {p^2 \over \Lambda^2}
+ \half (D-1) \thinspace {p^2 \over \Lambda^2}\Biggr\} \cr}$$
$$\eqalignno{&= -(D-1) \int_{R^2_1} {d^2\tau \over (\tau_1 + \tau_2)^{D/2}}
\thinspace \exp\Bigl(-{\tau_1 \thinspace \tau_2 \over \tau_1 +
\tau_2}\thinspace
{p^2 \over \Lambda^2}\Bigr) \Biggl\{{-\half D \over \tau_1 + \tau_2} + {\tau_1
\thinspace \tau_2 \over (\tau_1 + \tau_2)^2} \thinspace {p^2 \over \Lambda^2}
-\half {p^2\over \Lambda^2}\Biggr\} \cr
&- (D-2) \int_{R^2_1} {d^2\tau \over (\tau_1 + \tau_2)^{D/2}} \thinspace
\exp\Bigl(-{\tau_1 \thinspace \tau_2 \over \tau_1 + \tau_2}\thinspace {p^2
\over
\Lambda^2}\Bigr) \Biggl\{\half {(D-2) \over \tau_1 + \tau_2} + {\tau_1
\thinspace \tau_2 \over (\tau_1 + \tau_2)^2} \thinspace {p^2 \over \Lambda^2}
\Biggr\} &(4.11a) \cr}$$
$$\eqalignno{&= -(D-1) \int_{R^2_1} d^2\tau \enskip \half \Bigl({\partial \over
\partial \tau_1} + {\partial \over \partial \tau_2}\Bigr) \Biggl\{(\tau_1 +
\tau_2)^{-D/2} \thinspace \exp\Bigl(-{\tau_1 \thinspace \tau_2 \over \tau_1 +
\tau_2} \thinspace {p^2 \over \Lambda^2}\Bigr)\Biggr\} \cr
&+ \Bigl\{\int_0^1 dx \int_1^{\infty} dy + \int_1^{\infty} dx \int_{1/x}^{
\infty} dy\Bigr\} \thinspace {D-2 \over (x + 1)^{D/2+1}} {\partial \over
\partial y} \Biggl\{y^{1-D/2} \thinspace \exp\Bigl(-{xy \over x+1} \thinspace
{p^2 \over \Lambda^2}\Bigr)\Biggr\} \cr
&~ &(4.11b) \cr}$$
$$= 2 \Bigl({D-1 \over D-2}\Bigr) + \int_0^1 d\tau \enskip (\tau + 1)^{-D/2}
\thinspace \exp\Bigl(-{\tau \over \tau + 1} \thinspace {p^2 \over \Lambda^2}
\Bigr) \Biggl\{(D-1) - 2\Bigl({D-2 \over \tau + 1}\Bigr)\Biggr\} \eqno(4.11c)$$
These final terms serve to cancel the longitudinal contributions from $C$ and
$D$, proving transversality.

The integral identity we have just established can be used as well to simply
the
transverse part. The result is:
$$\Pi^{\alpha \beta}_{ab}(p) = {g^2 \over 2^D \thinspace \pi^{D/2}} \thinspace
f_{acd} \thinspace f_{bcd} \thinspace (p^2 \thinspace \eta^{\alpha \beta}
- p^{\alpha} \thinspace p^{\beta}) \thinspace \Pi(p^2) \eqno(4.12a)$$
$$\eqalignno{&\Pi(p^2) = \int_{R^2_1} d^2\tau {\Lambda^{D-4} \over (\tau_1 +
\tau_2)^{D/2}} \thinspace \exp\Bigl(-{\tau_1 \thinspace \tau_2 \over \tau_1
+ \tau_2} \thinspace {p^2 \over \Lambda^2}\Bigr) \Biggl\{2(D-2) {\tau_1
\thinspace \tau_2 \over (\tau_1 + \tau_2)^2} - \half (D-6)\Biggr\} \cr
&= 2 \int_0^{1/2} dx \enskip \Gamma\Bigl(2 - \frac{D}2, x{p^2 \over \Lambda^2}
\Bigr) \thinspace \Bigl[x (1-x) p^2\Bigr]^{{D\over 2}-2} \Bigl\{2(D-2)
\thinspace x \thinspace (1-x) - \half (D-6) \Bigr\} &(4.12b) \cr}$$
where the incomplete gamma function is:
$$\Gamma(n,z) \equiv \int_z^{\infty} dt \enskip t^{n-1} \thinspace e^{-t}
\eqno(4.13)$$
The dimensionally regulated result is obtained by the replacement:
$$\Gamma\Bigl(2-\frac{D}2,x {p^2\over \Lambda^2}\Bigr) \longrightarrow
\Gamma(2-\frac{D}2) \eqno(4.14)$$
This is an important check on accuracy. The same relation occurred for the one
loop vacuum polarization of nonlocal QED [7]; and the general rule relating the
two methods has been inferred for any diagram at any order in nonlocal scalar
field theory [13]. We conjecture the same rule applies for nonlocal Yang-Mills
as well. If true, this defines the general measure factor and the formalism is
complete.

Note that even if our conjecture is true nonlocal regularization would still
differ from the dimensional scheme. The latter regulates by deforming the
arguments of divergent Gamma functions while the former works by trading these
Gamma functions in for incomplete Gamma functions evaluated at predictable
functions of the Feynman parameters. One obvious distinction is that the
replacement (4.14) discards gauge invariant divergences for $D>4$. Dimensional
regularization is sensitive only to logarithmic divergences whereas nonlocal
regularization sees everything.

\vskip 1cm
\centerline{\bf 5. Acceptable nonlocalization as noncanonical quantization}

The first stage procedure of section 2 can be used to regulate any theory. The
result is basically a systematized version of proper time regularization. At
this level nonlocalization functions like any other regulator: the original
theory is changed to make it finite, but at the cost of introducing a
physically
unacceptable property which persists as long as the regulator is on. In order
to
obtain a physically acceptable theory one must take the unregulated limit after
the appropriate renormalizations have been performed.

For the case of nonlocal regularization the problem is that perturbative
unitarity is violated by loops which contain vector and tensor quanta. The
first stage contrives to enforce tree order decoupling through the agency of a
peculiar nonlinear gauge invariance. Since the naive functional measure does
not preserve this symmetry decoupling must fail in loop amplitudes. If we are
only interested in developing a noninvariant regulator this is as irrelevant as
the ghost states of the Pauli-Villars method; all will be set right in the
unregulated limit. However the situation changes radically if the second stage
can be carried out --- and we have seen some evidence that it can be for
Yang-Mills. In this case the regulated theory is not only finite but also
Poincar\'e invariant and perturbatively unitary. It has all the attributes of a
fundamental theory; why then must one take the unregulated limit?

One answer is that at fixed loop order our nonlocal amplitudes violate the
bounds implied by partial wave unitarity --- though of course the trees are
those of the local theory. This is perhaps not as damning as it seems because a
similar breakdown occurs even at tree order in string theory [15]. The problem
there has not been much studied but the feeling is that it can be resolved by
an ingenious resummation of perturbation theory [16]. Pending an argument
against the existence of the same mechanism in nonlocal field theory it is
obviously premature to fret overmuch about the violation of nonperturbative
bounds by finite order perturbation theory at extremely high energies.

The same considerations apply to the perturbative, off shell violations of
causality which appear as quantum effects in nonlocal field theory. Off shell
acausality occurs even at tree order in string field theory [8]. The effect has
received almost no attention because it does not give acausal poles in the
perturbative S-matrix, because it is extremely small (in nonlocal field theory
the strength of a signal propagated an invariant interval $\ell^2 >0$ outside
the lightcone would be suppressed by factors of $\exp[- \ell^2 \thinspace
\Lambda^2]$), and because essentially mystical considerations have encouraged
the belief that very small length scales cannot be probed in string theory.
In the rarified atmosphere of critical thinking which now prevails one might
advance any of these points with equal smugness as a reason for ignoring the
acausality of nonlocal field theory. Of greater import to us is the fact that
it
is only through {\it quantum effects} that the acausality of nonlocal field
theory manifests itself. If this acausality could be viewed as the consequence
of a noncanonical quantization procedure --- and we will shortly argue that it
can be --- then it might be both very small and logically consistent. We note
in
passing the observation of reference [8] that Planck scale acausality might
actually be {\it desirable} in providing an explanation of the horizon problem
of cosmology [17].

A more serious problem would seem to be the nonperturbative instabilities which
tend to occur in any fundamentally nonlocal theory. The origin of these
instabilities is explained at great length in reference [8]. Basically they are
due to the extra canonical degrees of freedom associated with higher time
derivatives. It is simple to show that if the Lagrangian contains up to $N$
time
derivatives then the associated Hamiltonian is linear in $N-1$ of the
corresponding canonical variables. A fully nonlocal Lagrangian whose
interactions are entire functions of the derivative operator can be viewed as
an
$N$th derivative Lagrangian in the limit that $N$ becomes infinite. Such a
nonlocal theory will then inherit the instabilities of its higher derivative
limit sequences {\it provided} the time dependence of the extra solutions does
not become arbitrarily choppy in the limit. The condition for the extra
solutions to have smooth limits is that no invertible field redefinition should
exist which takes the nonlocal field equations into local ones. That string
theory obeys this condition can be inferred from its tree order S-matrix; that
the nonlocal actions produced by the construction of section 2 {\it do not}
obey
this condition follows from the fact that their tree amplitudes agree with
those
of their local ancestors.

In fact we show in appendix A that the solutions of the nonlocal field
equations
are in one-to-one correspondence with those of the original theory. The
relation
is very simple:
$$\phi^{\rm nonloc.}_i = {\cal E}^2_{ij} \thinspace \phi^{\rm loc.}_j
\eqno(5.1)$$
Comparison of the actions shows that for solutions we also have $S[\phi] =
{\widehat S}[{\cal E}^2 \phi]$. {\it We stress that these are nonperturbative,
albeit classical, results. There are no extra classical solutions.} With no
extra solutions there is no necessary problem with either stability ---
although
the theory might still be unstable --- or with a finite initial value
formalism.

Since the on shell amplitudes of our nonlocalized action differ from those of
its local parent only at loop order it must be that the nonlocal field theory
can be viewed as a highly noncanonical quantization of fields which obey the
local equations of motion. The logical possibility of alternate quantization
schemes has long been realized but they have attracted little attention because
they tend not to result in unitary time evolution. {\it We stress that if the
second stage goes through then our construction does maintain unitarity, at
least perturbatively.}

We close with a discussion of another reason for the lack of interest in
noncanonical quantization, namely the spectacular phenomenological success
obtained by applying the canonical method to familar gauge theories. The
leftover interaction in all this is gravity. Indeed, one might regard the last
thirty years of effort in quantum gravity as a sort of proof by exhaustion of
a fundamental incompatibility between general relativity and quantum mechanics.
It is usually assumed that the former must be changed to accommodate the
latter;
we wish here to propose nonlocalization as a specific mechanism for achieving
the converse.

This is far too radical a proposal to be fully realized. For one thing it is
unaesthetically arbitrary; one can nonlocalize with any smearing operator
provided it is both analytic and convergent. Another problem is the background
dependence which seems to infiltrate the construction of section 2 from the
distinction it makes between the free and interacting parts of the action.
Finally, there is the issue of unification. Even if quantum gravity should owe
its ultraviolet finiteness to nonlocality we still expect it to form part of a
larger structure that includes the other forces in some way more fundamental
than as optional matter couplings.

These are all valid criticisms of the proposal --- and we have thought it good
to raise them in advance of anyone else. Nevertheless we have been unable to
dismiss the notion that certain forms of nonlocalization might, when applied to
special theories, yield a nonperturbatively acceptable theory. Nor have we been
able to rid ourselves of the feeling that noncanonical quantization, as
outlined
here, might play a decisive role in the quantum theory of gravitation.

\vskip 1cm
\centerline{\bf 6. Conclusions}

The major result of this paper is the procedure of section 2 for constructing a
regulated, nonlocal action ${\widehat S}[\phi]$ from any local action $S[\phi]$
which can be formulated perturbatively. The nonlocalized theory can either be
thought of as a regularization of the local theory or as a candidate
fundamental
model in its own right. Though ${\widehat S}[\phi]$ contains an infinite tower
of induced interactions these can be completely subsumed into an auxiliary
field
line which couples using only the local interaction. Loop calculations in this
formulation of the theory are no more difficult than with dimensional
regularization but the great thing about nonlocalization is its potential for
preserving symmetries. We prove in appendix B that ${\widehat S}[\phi]$
 preserves
suitably nonlocalized versions of any and all continuous symmetries of
$S[\phi]$.

Whether or not these symmetries are preserved in path integral quantization
depends upon the existence of an otherwise suitable measure factor $\mu[\phi]$
having the property that $[D\phi] \mu[\phi]$ is invariant. One must not expect
that all symmetries can be preserved in all theories; for example, gauge
invariance is broken in the chiral Schwinger model [12]. However, some
symmetries are certainly preserved in some theories; for example, the required
measure factor has been shown to exist for nonlocal QED [7]. No conclusion has
been reached yet for nonlocal Yang-Mills, although we did compute both the
invariant measure and the BRS measure at order $g^2$.

Quantization also requires that the gauge be fixed. A peculiarity of the
nonlocalized gauge symmetry is that its generators typically fail to close
except on shell. This means that the Faddeev-Popov determinant is not gauge
invariant, thus complicating the usual functional gauge fixing procedure. The
simplest solution to this problem is to nonlocalize the BRS theory directly and
then functionally quantize using a BRS invariant measure. When this is done one
finds both higher ghost interactions and higher ghost contamination in the
nonlocalized BRS transformation. Similar effects have been noted with
formulations of local supersymmetry which fail to close off shell [14].

Our partial determination of the measure factor was sufficient to evaluate the
vacuum polarization at one loop, which we did in section 4. As was the case
with
nonlocal QED [7], the result bears a striking resemblance to that obtained from
dimensional regularization. One merely replaces the divergent Gamma function of
the latter method with an incomplete Gamma function whose argument depends upon
the Schwinger parameters and ratios of momentum invariants to the
regularization
parameter $\Lambda^2$. For the vacuum polarization the necessary replacement is
given in equation (4.14). In another work we have presented the general
relation
for any amplitude at any order in scalar field theory [13]. If this same rule
applies as well to nonlocal Yang-Mills then the measure factor is determined.

The construction of section 2 can be applied to supersymmetric gauge theories
and to supergravity. Since it preserves the on shell tree amplitudes it will
necessarily preserve global supersymmetry (untouched since it is linearly
realized) and some sort of nonlocal and nonlinear generalization of gauge
invariance and local supersymmetry. If a supersymmetric measure factor can be
found as well then we will have produced the long-sought gauge and
supersymmetric invariant regulator.

Perhaps the most stimulating feature of our method is that it leaves unchanged
the tree amplitudes of the local theory. A consequence is that one ought to be
able to view nonlocal regularization as a noncanonical quantization of fields
obeying the local equations of motion. The connection can be made explicit and
we have sketched it in section 5. Though the idea is obviously incomplete at
this point, we feel it may play a role in reconciling gravitation and quantum
mechanics.

\vskip .5cm
\centerline{\bf Appendix A: Classical Solutions}

Here we prove a series of theorems concerning classical solutions to the
Euler-Lagrange equations associated with the local action, $S[\chi]$ --- for
the form of which see (2.1) --- the auxiliary action, ${\cal S}[\phi,\psi]$ ---
see (2.5) --- and the nonlocalized action, ${\widehat S}[\phi]$ --- see (2.7).

{\it Theorem A1: The shadow fields $\psi_i[\phi]$ can be expressed as follows:}
$$\psi_i[\phi] = -\Biggr({{\cal E}^2 - 1 \over {\cal E}^2}\Biggr)_{ij} \phi_j
+ {\cal O}_{ij} \thinspace {\delta {\widehat S}[\phi]\over \delta \phi_j}
\eqno(A.1)$$
To see this consider the variations of ${\cal S}[\phi,\psi]$:
$${\delta {\cal S}[\phi,\psi] \over \delta \phi_i} = {\cal E}^{-2}_{ij} {\cal
F}_{jk} \phi_k + {\delta I[\phi + \psi] \over \delta \phi_i} \eqno(A.2a)$$
$${\delta {\cal S}[\phi,\psi] \over \delta \psi_i} = - {\cal O}^{-1}_{ij}
\psi_j
+ {\delta I[\phi + \psi] \over \delta \phi_i} \eqno(A.2b)$$
Subtracting (A.2b) from (A.2a) and multiplying by the invertible operator
${\cal
O}$ gives:
$$\psi_i = -\Biggr({{\cal E}^2 - 1 \over {\cal E}^2}\Biggr)_{ij} \phi_j
+ {\cal O}_{ij} \thinspace \Biggr\{{\delta {\cal S}[\phi,\psi] \over \delta
\phi_j} - {\delta {\cal S}[\phi,\psi] \over \delta \psi_j}\Biggr\} \eqno(A.3)$$
We can neglect the rightmost term because $\psi_i[\phi]$ is defined to enforce
the vanishing of (A.2b). The result follows from the fact that at $\psi_i =
\psi_i[\phi]$ the variation of ${\widehat S}[\phi]$ is the same as (A.2a):
$$\eqalign{{\delta {\widehat S}[\phi] \over \delta \phi_i} &= {\delta {\cal S}
\over \delta \phi_i}\Bigl[\phi,\psi[\phi]\Bigr] + {\delta {\cal S} \over \delta
\psi_j}\Bigl[\phi,\psi[\phi]\Bigr] \thinspace {\delta \psi_j[\phi] \over \delta
\phi_i}\cr
&= {\delta {\cal S} \over \delta \phi_i}\Bigl[\phi,\psi[\phi]\Bigr] \cr}
\eqno(A.4)$$
The significance of this theorem is that $\psi_i[\phi]$ is unique up to terms
which vanish with the $\phi_i$ field equations. Since any such extra terms can
be absorbed into a field redefinition --- which only changes the still
undetermined measure factor --- it follows that the first stage of
nonlocalization is unique.

{\it Theorem A2: If the fields $\phi_i$ and $\psi_i$ obey the Euler-Lagrange
equations of ${\cal S}[\phi,\psi]$ then the field $\chi_i = \phi_i + \psi_i$
obeys the Euler-Lagrange equations of $S[\chi]$.} \hfill\break
\noindent Multiplying (A.2a) by ${\cal E}^2$, adding it to $(1 - {\cal E}^2)$
times (A.2b), and setting the sum to zero gives the equation:
$${\cal F}_{ij} (\phi_i + \psi_i) + {\delta I[\phi + \psi] \over \delta \phi_i}
= 0 \eqno(A.5)$$
Substituting $\chi_i = \phi_i + \psi_i$ proves the theorem.

{\it Theorem A3: If $\chi_i$ obeys the Euler-Lagrange equation of $S[\chi]$
then
the following fields:}
$$\phi_i = {\cal E}^2_{ij} \thinspace \chi_j \eqno(A.6a)$$
$$\psi_i = \Bigl(1 - {\cal E}^2\Bigr)_{ij} \thinspace \chi_j \eqno(A.6b)$$
{\it obey the Euler-Lagrange equations of ${\cal S}[\phi,\psi]$.} \hfill\break
\noindent By adding (A.6a) and (A.6b) we see that $\phi_i + \psi_i = \chi_i$.
Substitution into (A.2a) gives:
$${\delta {\cal S} \over \delta \phi_i}\Bigl[{\cal E}^2 \chi, (1 - {\cal E}^2)
\chi\Bigr]  = {\cal F}_{ij} \thinspace \chi_j + {\delta I[\chi] \over \delta
\chi_{i}} \eqno(A.7a)$$
Doing the same for (A.2b) gives:
$${\delta {\cal S} \over \delta \psi_i}\Bigl[{\cal E}^2 \chi, (1 - {\cal E}^2)
\chi\Bigr] = {\cal F}_{ij} \thinspace \chi_j + {\delta I[\chi] \over \delta
\chi_{i}} \eqno(A.7b)$$
The result follows from imposing the $\chi_i$ field equations.

{\it Theorem A4: If $\phi_i$ obeys the Euler-Lagrange equation of ${\widehat S}
[\phi]$ then $\chi_i = \phi_i + \psi_i[\phi]$ obeys the Euler-Lagrange equation
of $S[\chi]$.} \hfill\break
\noindent First note that $\phi_i$ and $\psi_i = \psi_i[\phi]$ enforce the
vanishing of (A.2a) --- by relation (A.4) --- and of (A.2b) --- from the
definition of $\psi_i[\phi]$. The result follows from theorem A2.

{\it Theorem A5: If $\chi_i$ obeys the Euler-Lagrange equation of $S[\chi]$
then
the field $\phi_i$, defined by (A.6a), obeys the Euler-Lagrange equation of
${\widehat S}[\phi]$.} \hfill\break
This result follows from (A.4) and theorem A3. Theorems A4 and A5 together
prove
the very important fact that the solutions of the nonlocalized Euler-Lagrange
equations are in one-to-one correspondence with those of the local action. It
follows that the nonlocalized theory is free of the sorts of higher derivative
solutions that wreck string field theory. Note also that the smearing operator
in (A.6a) generally smooths out small scale phenomena. Therefore it is entirely
possible for a singular solution of the local theory to give a nonsingular
solution in the nonlocalized theory.

\vskip .5cm
\centerline{\bf Appendix B: Classical Symmetries}

Here we prove a series of theorems concerning classical symmetries of the local
action, $S[\chi]$ --- for the form of which see (2.1) --- the auxiliary action,
${\cal S}[\phi,\psi]$ --- see (2.5) --- and the nonlocalized action, ${\widehat
S}[\phi]$ --- see (2.7).

{\it Theorem B1: If $S[\phi]$ is invariant under the infinitesimal
transformation:
$$\delta \phi_i = T_i[\phi] \eqno(B.1)$$
then the following transformation is a symmetry of ${\cal S}[\phi,\psi]$:}
$$\Delta \phi_i \equiv {\cal E}^2_{ij} \thinspace T_j[\phi + \psi]
\eqno(B.2a)$$
$$\Delta \psi_i \equiv (1 - {\cal E}^2)_{ij} \thinspace T_j[\phi + \psi]
\eqno(B.2b)$$
First note that by adding (B.2a) and (B.2b) we find:
$$\Delta (\phi + \psi)_i = T_i[\phi + \psi] \eqno(B.3)$$
Now from the definition of ${\cal S}[\phi,\psi]$ --- expression (2.5) --- it
follows that:
$$\eqalign{\Delta {\cal S}[\phi,\psi] &= \int d^Dx \enskip \Biggl\{\Bigl(\phi_i
+ \psi_i\Bigr) \thinspace {\cal F}_{ij} \thinspace T_j[\phi + \psi] + {\delta
I[\phi + \psi] \over \delta \phi_i} \thinspace T_i[\phi + \psi]\Biggr\} \cr
&= \Bigl(\delta S\Bigr)[\phi + \psi] \cr} \eqno(B.4)$$
Therefore we see that $\Delta {\cal S}[\phi,\psi] = 0$ as a consequence of
the assumed relation $\delta S[\phi] = 0$. A significant corollary is that
$[\delta_1,\delta_2] = \delta_3$ implies $[\Delta_1,\Delta_2] = \Delta_3$.

{\it Theorem B2: If ${\cal S}[\phi,\psi]$ is invariant under:
$$\Delta \phi_i = \tau_i[\phi,\psi] \eqno(B.5a)$$
$$\Delta \psi_i = \sigma_i[\phi,\psi] \eqno(B.5b)$$
then the following transformation is a symmetry of ${\widehat S}[\phi]$:}
$${\widehat \delta} \phi_i = \tau_i\Bigl[\phi,\psi[\phi]\Bigr] \eqno(B.6)$$
{}From definition (2.7) and the fact that $\psi_i[\phi]$ is constructed to
enforce
equation (A.2b) we see that:
$$\eqalign{{\widehat \delta} {\widehat S}[\phi] &= {\delta {\cal S}\Bigl[\phi,
\psi[\phi]\Bigr] \over \delta \phi_i} \thinspace {\widehat \delta} \phi_i +
{\delta {\cal S}\Bigl[\phi,\psi[\phi]\Bigr] \over \delta \psi_j} {\delta
\psi_j[\phi] \over \delta \phi_i} \thinspace {\widehat \delta} \phi_i \cr
&= \Bigl(\Delta {\cal S}\Bigr)\Bigl[\phi,\psi[\phi]\Bigr] \cr} \eqno(B.7)$$
Note that one generally obtains a different result from applying (B.6) to
$\psi_i[\phi]$ than by first transforming $\psi$ according to (B.5b) and then
setting $\psi_i = \psi_i[\phi]$:
$$\eqalignno{{\widehat \delta} \psi_i[\phi] &= {\delta \psi_i[\phi] \over
\delta
\phi_j} \thinspace \tau_j\Bigl[\phi,\psi[\phi]\Bigr] &(B.8a) \cr
&\neq \sigma_i\Bigl[\phi,\psi[\phi]\Bigr] &(B.8b) \cr}$$
The equality of (B.8a) and (B8.b) was unnecessary in proving theorem B2 because
$\psi_i[\phi]$ is defined to make the variation of ${\cal S}[\phi,\psi]$ with
respect to $\psi_i$ vanish. The same is not the case for closure: a simple
exercise shows that $[\Delta_1,\Delta_2] = \Delta_3$ implies $[{\widehat
\delta}_1, {\widehat \delta}_2] = {\widehat \delta}_3$ if and only if
${\widehat
\delta} \psi_i[\phi] = \sigma_i\Bigl[\phi,\psi[\phi]\Bigr]$.

{\it Theorem B3: If $S[\phi]$ is invariant under (B.1) then ${\widehat
S}[\phi]$
is invariant under:}
$${\widehat \delta} \phi_i = {\cal E}^2_{ij} \thinspace T_j\Bigl[\phi +
\psi[\phi]\Bigr] \eqno(B.9)$$
{\it The same transformation takes the shadow field to:}
$${\widehat \delta} \psi_i[\phi] = \Bigl(1 - {\cal E}^2\Bigr)_{ij} \thinspace
T_j\Bigl[\phi + \psi[\phi]\Bigr] - K_{ij}\Bigl[\phi + \psi[\phi]\Bigr]
\thinspace {\delta T_k \over \delta \phi_j}\Bigl[\phi + \psi[\phi]\Bigr]
\thinspace {\cal E}^2_{k\ell} \thinspace {\delta {\widehat S}[\phi] \over
\delta
\phi_{\ell}} \eqno(B.10a)$$
{\it where the operator $K_{ij}[\phi]$ is:}
$$K^{-1}_{ij}[\phi] \equiv {\cal O}^{-1}_{ij} - {\delta^2 I[\phi] \over \delta
\phi_i \delta \phi_j} \eqno(B.10b)$$
Of course relation (B.9) is a trivial consequence of the two preceding
theorems.
To prove (B.10) first apply the transformation to expression (A.1):
$$\eqalignno{{\widehat \delta} \psi_i[\phi] &= \Biggl({1 - {\cal E}^2 \over
{\cal E}^2}\Biggr)_{ij} \thinspace {\widehat \delta} \phi_j - {\cal O}_{ij}
\thinspace {\delta {\widehat \delta} \phi_k \over \delta \phi_j} \thinspace
{\delta {\widehat S}[\phi] \over \delta \phi_k} &(B.11a) \cr
&= \Bigl(1 - {\cal E}^2\Bigr)_{ij} \thinspace T_j\Bigl[\phi + \psi[\phi]\Bigr]
- {\cal O}_{ij} \thinspace \Biggl\{\delta_j^{~k} + {\delta \psi_k[\phi] \over
\delta \phi_j}\Biggr\} \thinspace {\delta T_{\ell} \over \delta \phi_k}\Bigl[
\phi + \psi[\phi]\Bigr] \thinspace {\cal E}^2_{\ell m} \thinspace {\delta
{\widehat S}[\phi] \over \delta \phi_m} \cr
&~ &(B.11b) \cr}$$
Now functionally differentiate the $\psi_i$ equation of motion:
$${\cal O}^{-1}_{ij} \thinspace {\delta \psi_j[\phi] \over \delta \phi_k} =
{\delta^2 I \over \delta \phi_i \delta \phi_j}\Bigl[ \phi + \psi[\phi]\Bigr]
\thinspace \Biggl\{\delta^j_{~k} + {\delta \psi_j[\phi] \over \delta \phi_k}
\Biggr\} \eqno(B.12a)$$
This can be rearranged to obtain an expression for ${\delta \psi_i}/{\delta
\phi_k}$ involving only $\phi$ and $\psi$:
$${\delta \psi_i[\phi] \over \delta \phi_k} = \Biggl\{ {\cal O}^{-1}_{ij} -
{\delta^2 I \over \delta \phi_i \delta \phi_j}\Bigl[\phi + \psi[\phi]\Bigr]
\Biggr\}^{-1} \thinspace {\delta^2 I \over \delta \phi_j \delta \phi_k}
\Bigl[\phi + \psi[\phi]\Bigr] \eqno(B.12b)$$
Substitution into (B.11b) and a few simplifications gives (B.10). A significant
corollary is that if the local symmetry obeys $[\delta_1, \delta_2] = \delta_3$
then the nonlocalized one obeys:
$$[{\widehat \delta}_1,{\widehat \delta}_2] \thinspace \phi_i = {\widehat
\delta}_3 \thinspace \phi_i + {\cal E}^2_{ij} \thinspace \Omega^{12}_{jk}\Bigl[
\phi + \psi[\phi]\Bigr] \thinspace {\cal E}^2_{k\ell} \thinspace {\delta
{\widehat S}[\phi] \over \delta \phi_{\ell}} \eqno(B.13a)$$
$$\Omega^{12}_{i\ell}[\phi] \equiv {\delta T^1_i[\phi] \over \delta \phi_j}
\thinspace K_{jk}[\phi] \thinspace {\delta T^2_{\ell}[\phi] \over \delta
\phi_k}
- {\delta T^2_i[\phi] \over \delta \phi_j} \thinspace K_{jk}[\phi] \thinspace
{\delta T^1_{\ell}[\phi] \over \delta \phi_k} \eqno(B.13b)$$
Therefore a local symmetry group nonlocalizes to a symmetry which must include
generators proportional to the field equations in order to give a closed
commutation algebra. As with so much else these generators are more easily
studied at the level of the auxiliary action.

{\it Theorem B4: The following transformation generates a symmetry of ${\cal S}
[\phi,\psi]$:}
$$\Delta \phi_i \equiv A_{ij}[\phi,\psi] \thinspace \Biggl\{{\delta {\cal S}[
\phi,\psi] \over \delta \phi_j} - {\delta {\cal S}[\phi,\psi] \over \delta
\psi_j} \Biggr\} \eqno(B.14a)$$
$$\Delta \psi_i \equiv - A_{ij}[\phi,\psi] \thinspace \Biggl\{{\delta {\cal S}[
\phi,\psi] \over \delta \phi_j} - {\delta {\cal S}[\phi,\psi] \over \delta
\psi_j} \Biggr\} \eqno(B.14b)$$
{\it provided $A_{ij}[\phi,\psi] = - A_{ji}[\phi,\psi]$.} \hfill\break
\noindent Of course $\Delta \psi_i = - \Delta \phi_i$. By simply
transforming the auxiliary action:
$$\eqalign{\Delta {\cal S}[\phi,\psi] &= \Bigl({\delta {\cal S} \over \delta
\phi_i} - {\delta {\cal S} \over \delta \psi_i}\Bigr) \thinspace \Delta
\phi_i \cr
&= \Bigl({\delta {\cal S} \over \delta \phi_i} - {\delta {\cal S} \over \delta
\psi_i}\Bigr) \thinspace A_{ij}[\phi,\psi] \thinspace \Bigl({\delta {\cal S}
\over \delta \phi_j} - {\delta {\cal S} \over \delta \psi_j}\Bigr) \cr}
\eqno(B.15)$$
we see that the theorem follows trivially from the antisymmetry of $A_{ij}$.
Since $\psi_i[\phi]$ is defined to make ${\delta {\cal S}}/{\delta \psi_i} = 0$
the image of this transformation under theorem B2 gives the most general off
shell symmetry of ${\widehat S}[\phi]$. Such symmetries occur in all field
theories --- even conventional, local ones. Of course generators which have no
effect upon solutions of the field equations are without dynamical
significance. We consider them here only because some subset of these
symmetries
must be added to the dynamically significant generators in order to obtain a
closed commutation algebra.

Comparison with relation (A.2) reveals that transformation (B.14a) can be
recast
in the following simple form:
$$\Delta \phi_i = A_{ij}[\phi,\psi] \thinspace {\cal E}^{-2}_{jk} \thinspace
{\cal O}^{-1}_{k \ell} \thinspace \Bigl[\Bigl(1 - {\cal E}^2\Bigr)_{\ell m}
\thinspace \phi_m - {\cal E}^2_{\ell m} \thinspace \psi_m\Bigr] \eqno(B.16)$$
An important special case is given by the choice:
$$A_{ij}[\phi,\psi] = M_{ij} \thinspace {\cal O}_{jk} \thinspace {\cal E}^2_{k
\ell} \eqno(B.17)$$
where $M_{ij}$ is an antisymmetric operator which commutes with ${\cal
F}_{ij}$.
When $\delta$ is the BRS symmetry the nonlocalized transformation rule induced
by theorems B1 and B2 turns out not to agree with the local rule even at lowest
order. Instead the lowest order terms differ by a factor of ${\cal E}^2$. We
have found it convenient to absorb this factor by using the image of one of the
dynamically trivial symmetries of type (B.17).

\vskip .5cm
\centerline{ACKNOWLEDGEMENTS}

We have benefitted from conversations with M. Chu, A. Polychronakos, P. Ramond
and D. Zoller. This work was partially supported by the Institute for
Fundamental Theory and by DOE contracts DE-FG05-86-ER40272 and AS05-80ER-10713.

\references
\doublespace

[1] A. M. Polyakov, {\sl Phys. Lett.} {\bf 103B} (1981) 207; \hfill\break
D. Friedan, in {\it Recent Advances in Field Theory and Statistical Mechanics},
eds. J.-B. Zuber and R. Stora (North Holland, Amsterdam, 1984); \hfill\break
O. Alvarez, {\sl Nucl. Phys.} {\bf B216} (1983) 125.

[2] S. B. Giddings, {\sl Nucl. Phys.} {\bf B278} (1986) 242.

[3] D. J. Gross and A. Jevicki, {\sl Nucl. Phys.} {\bf B283} (1987) 1;
\hfill\break E. Cremmer, A. Schwimmer and C. B. Thorn, {\sl Phys. Lett} {179B}
(1986) 468.

[4] C. Hayashi, {\sl Prog. Theor. Phys.} {\bf 10} (1953) 533; {\bf 11} (1954)
226; \hfill\break
R. Marnelius, {\sl Phys. Rev.} {\bf D8} (1973) 2472; {\bf D10} (1974) 3411.

[5] J. Polchinski, {\sl Nucl. Phys.} {\bf B231} (1984) 269.

[6] X. Ja\'en, J. Llosa and A. Molina, {\sl Phys. Rev.} {\bf D34} (1986) 2302.

[7] D. Evens, J. W. Moffat, G. Kleppe and R. P. Woodard, {\sl Phys. Rev.} {\bf
D43} (1991) 499.

[8] D. A. Eliezer and R. P. Woodard, {\sl Nucl. Phys.} {\bf B325} (1989) 389.

[9] B. W. Lee and J. Zinn-Justin, {\sl Phys. Rev.} {\bf D5} (1972) 3121.

[10] L. D. Faddeev and A. A. Slavnov, {\it Gauge Fields: Introduction to
Quantum
Field Theory} (Benjamin/Cummings, Reading, 1980).

[11] Z. Bern, M. B. Halpern, L. Sadun and C. Taubes, {\sl Phys. Lett.} {\bf
B165} (1985) 151; {\sl Nucl. Phys.} {\bf B284} (1987) 1; {\sl Nucl. Phys.} {\bf
B284} (1987) 35.

[12] B. Hand, ``Nonlocal Regularization of the Chiral Schwinger Model,''
University of Toronto preprint. (To appear in {\sl Physics Letters} {\bf B}.)

[13] G. Kleppe and R. P. Woodard, ``Two Loop Calculations Using Nonlocal
Regularization,'' University of Florida preprint UFIFT-HEP-91-20, November
1991.

[14] B. de Wit and J. W. van Holten, {\sl Phys. Lett.} {\bf 79B} (1978) 389.

[15] M. Soldate, {\sl Phys. Lett.} {\bf 186B} (1987) 321.

[16] I. J. Muzinich and M. Soldate, {\sl Phys. Rev.} {\bf D37} (1988) 359.

[17] A. H. Guth, {\sl Phys. Rev.} {\bf D23} (1981) 347.

\endreferences

\vfill\eject
\centerline{FIGURE CAPTIONS}
\doublespace

\noindent Fig. 1a: The smeared and shadow gluon propagators of nonlocal
Yang-Mills.

\noindent Fig. 1b: The smeared and shadow ghost propagators of nonlocal
Yang-Mills.

\noindent Fig. 2a: The 2-point measure vertex of nonlocal Yang-Mills.

\noindent Fig. 2b: The 3-point vertices of local Yang-Mills.

\noindent Fig. 2c: The 4-point vertex of local Yang-Mills.

\noindent Fig. 3: The component $A^{\alpha \beta}_{ab}(p)$ of the one loop
vacuum polarization.

\noindent Fig. 4: The component $B^{\alpha \beta}_{ab}(p)$ of the one loop
vacuum polarization.

\noindent Fig. 5: The component $C^{\alpha \beta}_{ab}(p)$ of the one loop
vacuum polarization.

\noindent Fig. 6: The component $D^{\alpha \beta}_{ab}(p)$ of the one loop
vacuum polarization.

\bye